# Halide perovskites: third generation photovoltaic materials empowered by metavalent bonding

Matthias Wuttig[1,2,3], Carl-Friedrich Schön[1], Mathias Schumacher[4], John Robertson[5], Pavlo Golub[6], Eric Bousquet[7], Jean-Yves Raty[8,9]

1) I. Institute of Physics, Physics of Novel Materials, RWTH Aachen University, 52056 Aachen, Germany
2) Jülich-Aachen Research Alliance (JARA FIT and JARA HPC), RWTH Aachen University, 52056 Aachen, Germany
3) PGI 10 (Green IT), Forschungszentrum Jülich GmbH, 52428 Jülich, Germany
4) Institute for Theoretical Solid State Physics, RWTH Aachen University, 52056 Aachen, Germany
5) Engineering Dept, Cambridge University, Cambridge CB3 0FA, UK
6) J. Heyrovský Institute of Physical Chemistry, Department of Theoretical Chemistry, Dolejškova 2155/3, 182 23 Prague 8, Czech Republic
7) Physique Théorique des Matériaux, Q-MAT, CESAM, Université de Liège, B4000 Sart-Tilman, Belgium
8) CESAM and Physics of Solids, Interfaces and Nanostructures, B5, Université de Liège, B4000 Sart-Tilman, Belgium
9) UGA, CEA-LETI, MINATEC campus, 17 rue des Martyrs, F38054 Grenoble Cedex 9, France

**Abstract:**

Third-generation photovoltaic (PV) materials combine many advantageous properties, including a high optical absorption together with a large charge carrier mobility, facilitated by small effective masses. Halide perovskites ($ABX_3$, where X = I, Br or Cl) appear to be the most promising third-generation PV materials at present. Their opto-electronic properties are governed by the B-X bond. A quantum-chemical bond analysis reveals that this bond differs significantly from ionic, metallic or covalent bonds. Instead, it is better regarded as metavalent, since it shares approximately one p-electron between adjacent atoms. The resulting σ–bond is half-filled, which causes pronounced optical absorption. Electron transfer and lattice distortions open a moderate band gap, resulting in charge carriers with small effective masses. Hence metavalent bonding explains the favorable PV properties of halide perovskites. This is summarized in a map for different bond types, which provides a blueprint to design third-generation PV materials.

Halide perovskites promise exceptional performance in optoelectronic applications ranging from inexpensive, high-performance photovoltaic (PV) modules (*1-5*) to light- emitting and lasing devices (*6-8*). These perovskites display a rare combination of properties including pronounced optical absorption in conjunction with relatively large charge carrier mobilities. Yet, they possess soft crystalline lattices with dynamic disorder (*9*). Such a combination of properties is neither found in isostructural oxide perovskites nor sp$^3$-bonded semiconductors. While the remarkable properties of halide perovskites enable their application potential (*10-12*), their origin is not yet fully understood (*9*). Here we demonstrate that the unique property portfolio of halide perovskites is empowered by an unconventional bonding mechanism not observed in other prominent PV materials or isostructural oxide perovskites. This insight provides a blueprint for the design of tailored halide perovskites and related materials for PV applications.

Perovskites form a wide class of versatile compounds where nearly all atoms of the Periodic Table can fit into its chemical formula $ABX_3$. The most studied perovskites were for a long time the oxides (X = O) due to their numerous applications including sonars (piezoelectric $PbZr_xTi_{1-x}O_3$ - PZT) (*13*), photonics (electro-optic $LiNbO_3$) (*14*), capacitors (dielectric permittivity, BaTiO3) (*15*), and infrared detectors (pyroelectric $LiTaO_3$, $PbTiO_3$) (*16*). Only very recently a surge of interest has been stirred focusing on halide perovskites and their potential for opto-electronic devices, most notably at present for PV applications (*1-5*). These halide perovskites are characterized by high charge carrier mobilities and have an imaginary dielectric function $\varepsilon(\omega)$ that maximizes absorption in the visible range, with a steep raise just after the absorption edge (the most important properties are listed in **Tab. 1**).

| | |
|---|---|
| **Optical properties** | **Direct band gap of 1.1 – 1.4 eV** |
| | sharp and steep absorption edge |
| | low non-radiative recombination rate |
| **Electrical properties** | **Charge carriers with high mobility $\mu$** |
| | defect scattering small |
| **Sample preparation** | **Ease of preparation of high quality samples** |
| | low temperature preparation route |
| | inexpensive and abundant elements |

**Table 1: Desired material properties for the absorber material in a solar cell.** Similar requirements can be listed for other opto-electronic devices such as photo-diodes or solid state lasers.

To derive a rational design rule for PV materials it is crucial to understand the key microscopic parameters responsible for these unique PV properties of halide perovskites. To that end, we compare halide perovskites with their oxide counterparts. This evaluation reveals that halide perovskites employ an unconventional bonding mechanism, which is the key for their extraordinary combination of properties. To make this assessment, we will first focus on the cubic, inorganic perovskites. Subsequently, a comparison of inorganic and hybrid organic-inorganic halide perovskites is presented.

**Quantum-Chemical Analysis of Bonding in Halide Perovskites**

A systematic understanding of bonding can be obtained by quantum chemical tools such as the quantum theory of atoms in molecules (QTAIM) (*17*). With this theory, which goes beyond the one electron density functional theory, two quantities can be determined, which quantify chemical bonding: the electron transfer and the number of electrons shared (ES) between pairs of neighbouring atoms (see methods section) (*18-20*). To facilitate a comparison of different compounds, the relative electron transfer (ET) is utilized, which is obtained upon dividing the total electron transfer by the oxidation state (nominal ionic charge). An electron sharing between neighbouring atoms equal to 2 corresponds to the Lewis picture of a perfect covalent bond, i.e. an electron pair. ET and ES are accurate quantum-chemical bond indicators, which can separate the different types of bonding as shown in **Figure 1**. Materials which employ ionic bonding are located in the lower right corner of the map, since they are characterized by a significant electron transfer. In covalent compounds, on the contrary, up to 2 electrons (an electron pair) are shared between neighbouring atoms (upper left corner). We can now explore how the bonding in perovskites compares with metallic, ionic and covalent bonding, and hence where the bonds in the different perovskites are located on the map.

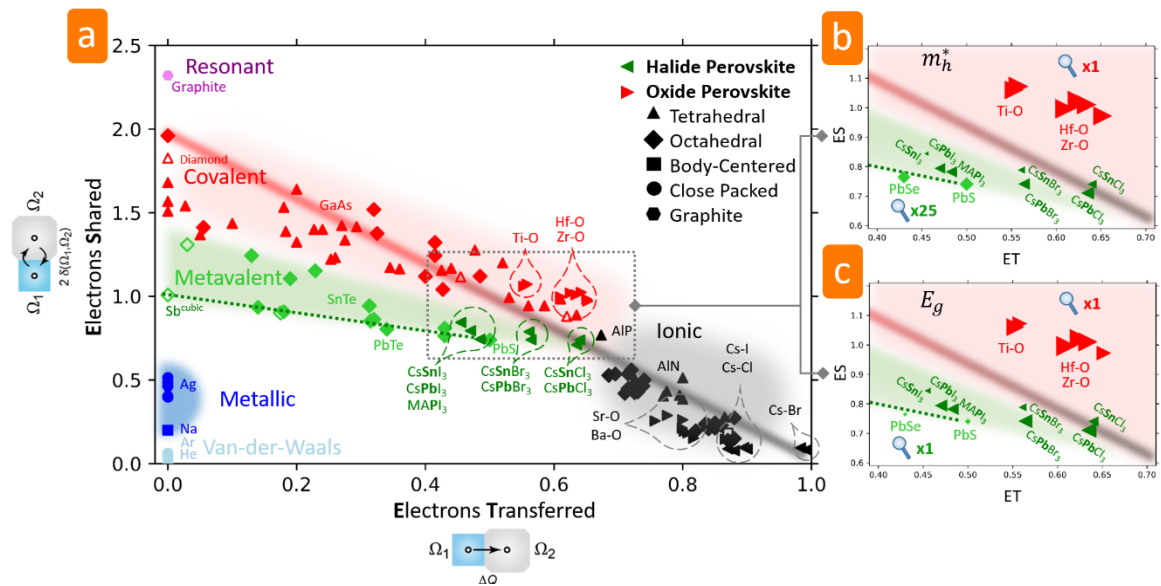

**Figure 1: a) 2D map describing the bonding in solids.** The map is spanned by the number of electrons shared between adjacent atoms and the electron transfer renormalized by the oxidation state. Different colors characterize different material properties (*21*) and have been related to different types of bonds (*22*). Triangles pointing left are utilized for halide perovskites, those pointing right for oxide perovskites. For these compounds, the element in bold characterizes the bond considered. Interestingly, the bonds for the different perovskites fall into three distinctively different classes. The red – black line describes the transition from covalent bonds to ionic bonds. The dashed green line denotes metavalently bonded solids with perfect octahedral arrangement. MAPI stands for $CH_3NH_3PbI_3$. **b) and c) Hole effective masses $m_h^*$ and band gaps $E_g$ for different perovskites** as a function of chemical bonding, where the symbol size represents the size of $m_h^*$ and $E_g$. The hole effective masses of the halide perovskites, which have been magnified by a factor of 25 for the sake of comparison, are much smaller than those for the oxide perovskites. The $m_h^*$ values of the halide perovskites are closely related to the size of the band gap and increase with increasing electron transfer.

The majority of the compounds depicted in the map are either elements or binary compounds of A-B type, where a single type of bond is sufficient to brand the solid. In ABX$_3$ perovskites, on the contrary, we have to describe the A-X and B-X bond separately, since they will be characterized by different values for the number of electrons shared and transferred. This enables us to relate different properties to these two different bonds and will help to unravel the structure – property relationship based on an in-depth understanding of chemical bonding. In the following we will focus first on undistorted perovskites, i.e. perfectly cubic systems with $Pm\bar{3}m$ symmetry. By comparing halide and oxide based perovskites which share the same crystal structure, we can also explore which material properties are governed by differences in bonding. The resulting data are summarized in **Fig. 1**.

**Figure 1** reveals that the A-X bond is always ionic, both for the oxide and the halide perovskites. On the contrary, the B-X bonds differ between the two perovskite families. The assignment of a bond as being ionic or covalent is accomplished based on the corresponding material properties (*21*). To distinguish between ionic and covalent bonds, the coordination number and the chemical bond polarizability are employed. In the supplement, it is shown that the A-X bond in oxide and halide perovskites closely resemble each other and are identified as ionic bonds. The B-O bond in the oxides, on the contrary, is best described as iono-covalent. There is both considerable charge transfer and significant sharing of electrons, i.e. a combination of significant ionic and covalent bonding contributions. Here, we focus on the B-X bond, which will be shown to be responsible for the PV properties of the halide perovskites. Interestingly, as shown in **Figure 1**, the B-X bond in the halide perovskites is located in a distinct, neighboring region. While there is significant electron transfer, ET is considerably smaller for the B-X than the A-X bond. Yet, it is also not a typical iono-covalent bond either. The corresponding ES values are clearly below those of the B-O bonds and they are also much more polarizable (see **table S1** and **S2**). Furthermore, for a covalent halogen bond, one expects an effective coordination number of 1, as observed in fluorocarbons, for example. The coordination number of 2 for the halogen atoms is hence incompatible with a simple covalent bond. Instead, the ES values resemble those found in metavalently bonded compounds such as PbTe or PbSe (*23*). This is no coincidence but instead indicative for a common type of bonding invoked in both halide perovskites and lead chalcogenides as will be shown below.

**From unconventional Bonding to favorable Material Properties**

So far, it has been argued that there are differences between oxide and halide perovskites concerning the B-X bond. Can those differences explain and help to design material properties which are decisive for PV applications? A crucial property for PV materials is the effective mass of the charge carriers in the vicinity of the Fermi level. This quantity shows striking differences between oxide and halide perovskites as depicted in **Figure 1.b)** and **S1**. In the oxide perovskites, both the electron and the hole effective mass, i.e. $m_e^*$ and $m_h^*$ are significantly larger than the free electron mass $m_e$. Such high effective masses are indicative for low mobilities, a major disadvantage for opto-electronic applications. The halide perovskites, on the contrary, have much lower effective masses. The hole effective mass $m_h^*$ is even significantly lower than the free electron mass (*24*) and shows a clear chemical trend. The replacement of I by Br and even more so Cl increases the hole effective mass. The combination of strong optical absorption (discussed below) and small effective masses are ideal for PV applications and identify halide perovskites as prototypes of third generation PV materials.

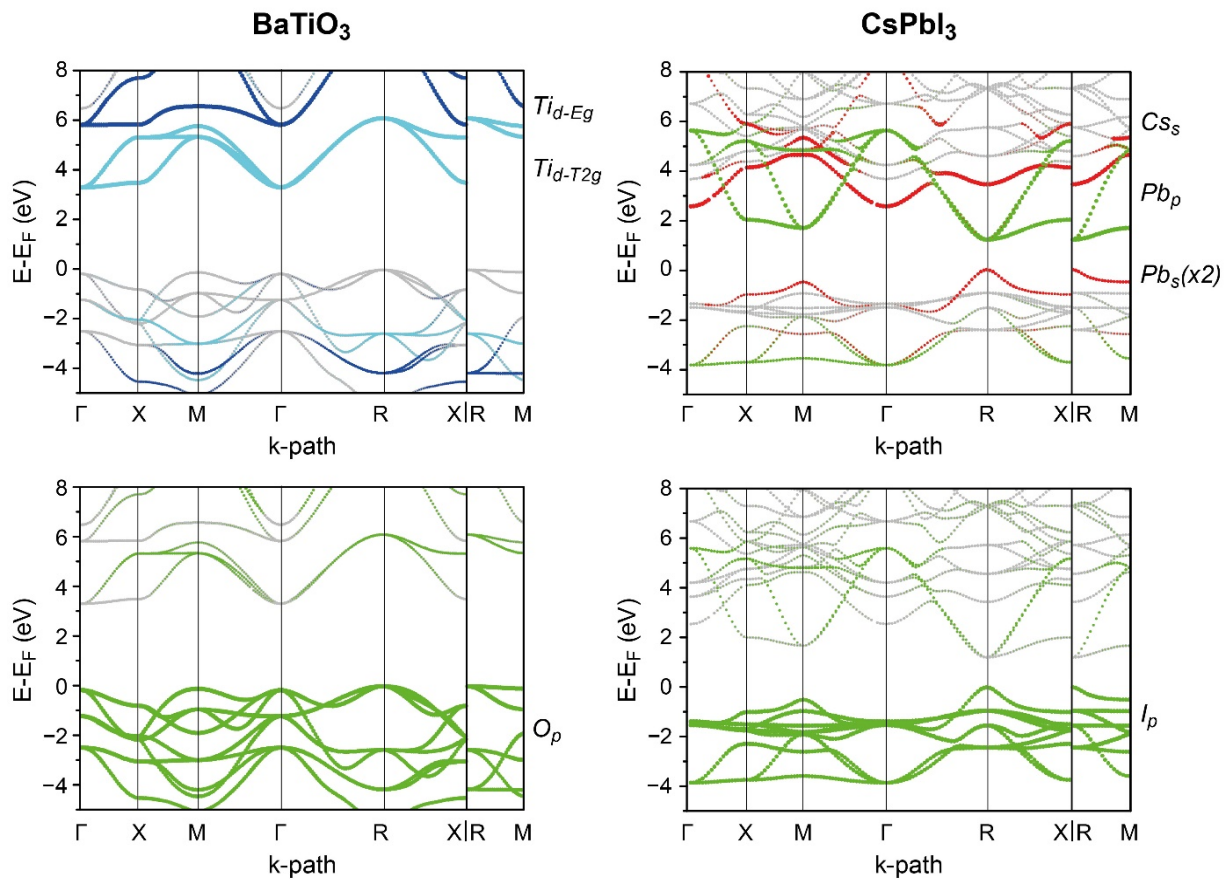

**Figure 2**: **Band structure decomposition of two characteristic perovskites, BaTiO$_3$, left, and CsPbI$_3$, right.** s-, p- and d-states are plotted in red, green and blue, respectively and the area of the symbols is proportional to the orbital contribution to the bands. Top panels: Fat bands for Ti d$_{xy}$, d$_{yz}$, d$_{xz}$ (cyan, labeled Ti$_{d-T2g}$), and Ti d$_{x2-y2}$ and d$_{3z2-r2}$ (blue, label Ti$_{d-Eg}$) in BaTiO$_3$ and Pb-p, Pb-s and Cs-s in CsPbI$_3$. To improve the visibility of the Pb-s contribution, its magnitude has been increased by a factor 2. Bottom panels: Fat bands for O-p in BaTiO$_3$ and I-p in CsPbI$_3$. The complete band structure is given in **Figure S2**.

These differences between halide and oxide perovskites and the dependence of $m_h^*$ on stoichiometry are indicative for systematic changes in the band structure, which reflect related differences in chemical bonding (see also **Figure 1.c**). Both optical properties and effective masses of the charge carriers are intimately interwoven with the character of the states and the band structure in the vicinity of the Fermi energy, depicted in **Figure 2**. The valence band (VB) structure presents many similarities between the two perovskite families, the main common feature being the location of the VB maximum at the R-point. This comes from the dominant role of the p-orbitals of the X atom in these bands. Yet, the conduction band minimum is located at different points in reciprocal space. This is due to the different nature of the states in the conduction bands. For the halide perovskites the relevant states are predominantly p-states of the B atom, while for the oxide perovskites the d-states of the B atom (here: Ti) are important.

To understand the relationship between chemical bonding, band structure and PV properties, we analyze the atomic orbitals responsible for the bond formation. These orbitals are listed in **Table S3**. For both Pb and I, the orbitals in the vicinity of the Fermi level are crucial. As depicted in **Figure 3** on the left hand side (and in **Table S3**), both for Pb and I atoms, the outermost atomic orbitals are p-orbitals. In the solid (center of **Fig. 3**), their overlap is shown, leading to the formation of a σ-bond. The actual number of shared electron pairs is determined by the quantum-chemical calculations (**Tab. S3 - S6**). Between adjacent Pb and I atoms a σ-bond is formed, which contains 0.8 electrons shared between the two atoms (see **Fig. 1**), in line with a standard analysis of the DFT data (**Tab. S8**). Slightly lower values are obtained for the Pb-Br bond in CsPbBr$_3$ and the Pb-Cl bond in CsPbCl$_3$. Hence, with about one electron (a half-filled orbital) between adjacent Pb and I atoms and significant overlap of the outermost atomic orbitals, we are expecting a metallic band, as shown on the right side of **Figure 3**. Halide perovskites are hence incipient metals (*21*), i.e. have an almost metallic ground state.

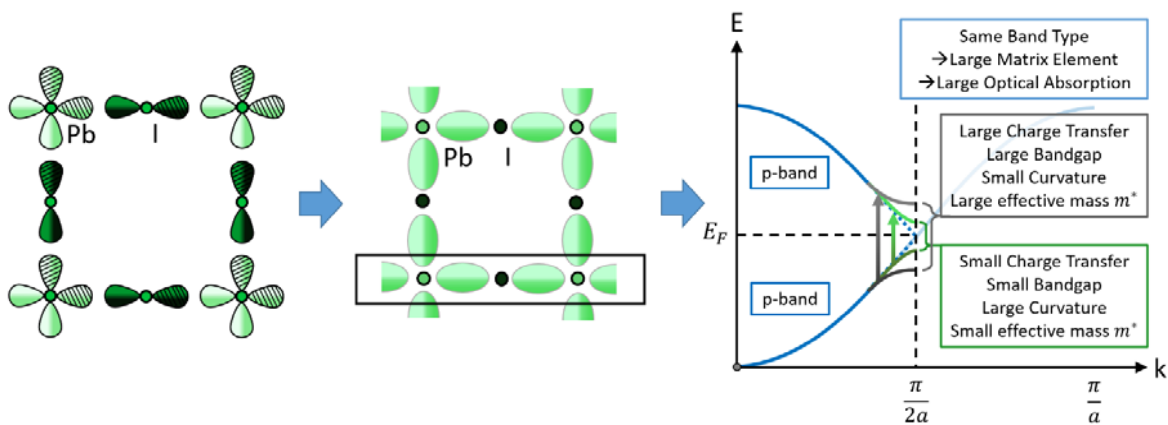

**Figure 3: Bond formation and resulting band structure in CsPbI$_3$.** Atomic p-orbitals of Pb and I responsible for bond formation are depicted on the left hand side. The atomic arrangement in the (001) plane is shown, characterized by small electron transfer and the formation of σ-bonds between adjacent atoms. These σ-bonds are occupied by about half an electron pair (ES≈ 1), resulting in a metallic band (blue curves on the right side). However, electron transfer creates a band gap.

Yet, there are two mechanisms, which will open a band gap, a necessity for PV materials. These are either distortions of the linear alignment for the B – X – B chain (Peierls' distortions) and/or charge transfer. This opening of the band gap also has a strong impact on the charge carriers. The sketch in **Figure 3** already reveals that an increasing band gap leads to a larger effective mass, in line with the data, which show a clear correlation between $E_g$ and $m_h^*$ (see **Fig. 1.b) and c)**). Interestingly, a simple tight binding model reproduces these claims (section S.IV). Consider a chain of I-p and Pb-p orbitals with distance $a$ and onsite energies ($\varepsilon_{I-p}$, $\varepsilon_{Pb-p}$), as well as resonance integral $\beta_{pp}$ between them, setting $\varepsilon_{I-p} - \varepsilon_{Pb-p} = 2\Delta$. In this case, one obtains $E_G = 2\Delta$, while the effective mass at R equals $\pm \frac{1}{\hbar^2} \frac{\Delta}{a^2 \beta^2}$. A decrease of charge transfer thus leads to smaller effective masses. The nature of the B-X bonds, its metavalence, is hence responsible for the favorably small effective masses of the halide perovskites. One can even explain the asymmetry of the effective masses depicted in **Figure S1**. This is due to a small contribution of Pb-s orbitals at the top of the valence band, which simultaneously decreases $m_h^*$ but increases $m_e^*$ (section S.IV).

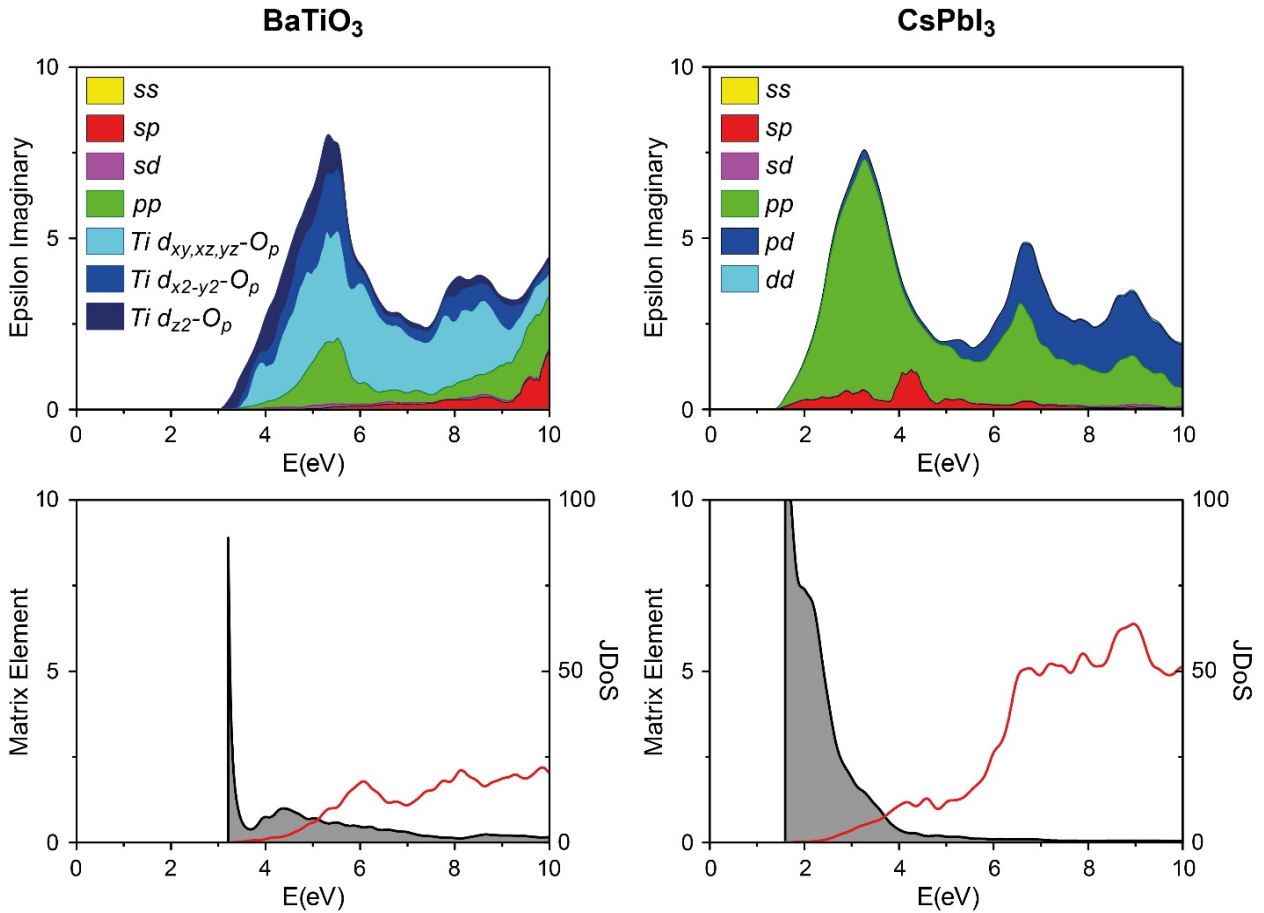

**Figure 4**: **Absorption (epsilon imaginary) for BaTiO$_3$ and CsPbI$_3$ (top), decomposed into the joint density of states and the corresponding average matrix element (bottom).** In CsPbI$_3$, the absorption maximum is observed at lower energy. It benefits from a large matrix element, dominated by p – p transitions. The optical properties of halide perovskites are thus far superior for PV applications.

With this information, the optical properties of halide perovskites can be analyzed, too. The imaginary part of the dielectric function characterizes the absorption of photons of energy $\hbar\omega$. As depicted in **Figure 4**, $\varepsilon_2(\omega)$ increases steeply above the band gap and reaches a pronounced maximum for both perovskites. The band gap for solar cell materials should be between 1.1 and 1.4 eV, according to Shockley and Queisser (*25*). This can be realized easily in the halide perovskites listed in **Table S2**. For oxide perovskites, $BaTiO_3$ has a larger band gap of 3.18eV, and the maximum absorption only occurs at 5.31 eV, too high for PV applications. Yet, there is another important difference. A comparison of the maximum of $\varepsilon_2(\omega)$ shows similarly high values, even though the joint density of states is higher for the oxide perovskites. This is explained by a larger transition matrix element for the halide perovskites, attributed to the nature of the states involved. For the oxide perovskites, the oxygen p – metal d- transition governs the absorption, while the oxygen s- – metal d- transition hardly contributes. The halide perovskites, on the contrary, reveal a strong peak for I p- to Pb p-state transitions. Utilizing a transition between states of same character, i.e. p-states, yet opposite parity is ideal, ensuring pronounced orbital overlap and hence strong absorption. The σ–bond formed from adjacent p-states of B and X atoms in conjunction with the moderate band gap thus explains both the strong optical absorption and the small effective masses simultaneously. It should be noted that possible structural distortions have to remain moderate to preserve metavalence. Indeed, several studies of chalcogenide phase change materials pointed to the fact that p-bonding can evolve towards covalence with the appearance of disorder, accounting for the electrical and optical contrast (*26, 27*).

Before discussing how the insights obtained can be utilized to tailor materials for PV applications, we discuss in the supplement how the transition from inorganic halide perovskites to hybrid organic – inorganic halide perovskites will alter PV material properties. Furthermore, the impact of distortions away from a cubic atomic arrangement will be evaluated. Both the impact of an organic molecule at the A-site and distortions of the B atom away from a perfect octahedral arrangement can be described in the framework presented here.

**Conclusions and Outlook**

As we have shown, the bonding configuration in halide perovskites is distinctively different from metallic, ionic and covalent bonding. The unconventional bonding mechanism utilized in these perovskites explains many of their favorable PV characteristics. One should thus expect similarly attractive properties in other materials which employ this type of bonding. As shown in section S.IX, the band structures of lead chalcogenides closely resemble those of halide perovskites. In both cases the p-states in the vicinity of the Fermi level form half-filled σ-bonds, leading to small effective masses and a strong optical absorption. Interestingly, the optical absorption is even significantly higher for lead chalcogenides since the chalcogen and lead atom form 6 σ-bonds each, while a halogen atom only forms two σ-bonds in the perovskites. This comparison also helps understanding another important aspect of charge transport. While both halide perovskites and lead telluride have attractively small effective masses, they only possess moderate mobilities at room temperature. This is due to strong electron-phonon coupling, responsible for the anomalously large Born effective charge. While a higher mobility would further improve the performance of these for PV materials, the strong electron-phonon coupling does not trap the charge carriers in contrast to deep defects or polarons. Therefore, the carrier's lifetime is longer, which ensures a high mobility - life-time product, a key factor for the efficiency of these PV materials.

The quantum mechanical indicators computed also allow to extract important trends that can help designing future PV materials. **Figure 5** shows for instance, how the band gap varies with the number of shared and transferred electrons. Among all compounds for which the data have been computed, the halide perovskites occupy a well-defined area of the metavalent region of the map.

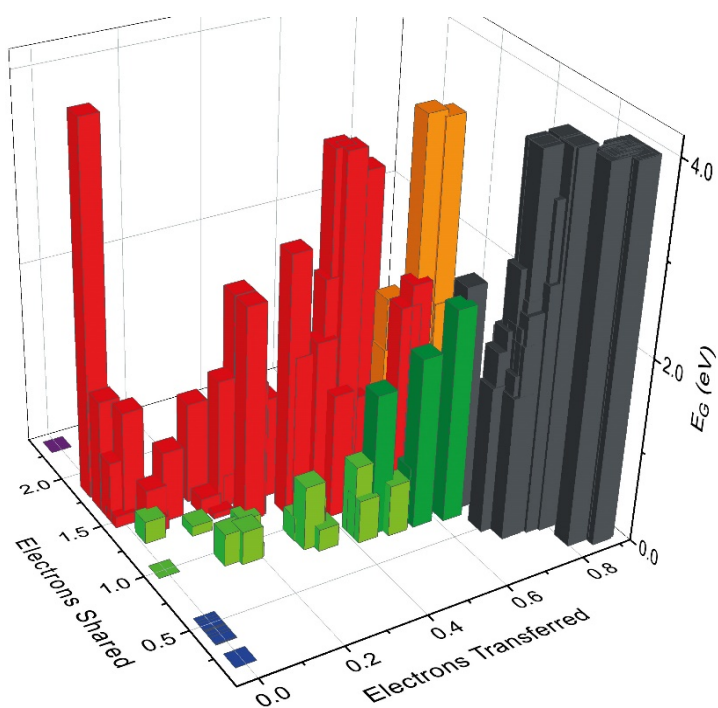

**Figure 5**: **Band gap as a function of electrons transferred and shared.** Different types of bonds are denoted by different colors, i.e. ionic (black), covalent (red), metallic (blue), metavalent (green), halide perovskites (light green) and oxide perovskites (orange). Metavalently bonded systems like PbTe or $CsSnI_3$ are characterized by sharing about one electron between adjacent atoms and modest charge transfer, leading to small band gaps, strong optical absorption and small effective masses. The number of electrons shared and transferred can be employed to tune the band gap and related PV properties.

The compounds we have computed are characterized by an octahedral arrangement of the B atom. These halide perovskites are all located in the vicinity of the dashed line in **Figure 1**, where we also find the octahedrally coordinated lead chalcogenides. As can be seen from a comparison of $CsSnI_3$, $CsSnBr_3$ and CsSnCl$_3$, their band gap is largely controlled by the electron transfer. Increasing the electron transfer, or alternatively the number of electrons shared, while maintaining metavalent bonding increases the band gap, yet preserves favorably small effective masses and strong optical absorption. Another option to fine-tune the optical properties is to tailor the interatomic distance. Both for gaps created by charge transfer (halide perovskites, lead chalcogenides) and gaps created by Peierls-distortions, expanding the metavalent σ–bond increases the gap. Therefore, one can e.g. modify the A atom in ABX$_3$ halide perovskites to stretch the B-X bond. The map in **Figure 1** hence provides a blueprint to design 3$^{rd}$ generation PV materials.

**Methods:**

All structure and electronic structure calculations were performed using Density Functional Theory in combination with projector augmented wave (*28*) (PAW) potentials as implemented in the ABINIT (*29, 30*) and VASP (*Vienna ab initio simulation package*) codes (*31-35*) . Different exchange-correlation functionals were used. The SCAN (Strongly Constrained and Appropriately Normed semilocal density functional) meta-GGA functional (*36*) was used together with the VASP code to compute ground state electronic properties. This functional has been shown to outperform common local functionals (*37*). In particular, it yields gap values that are closer to the experimental ones, although these remain underestimated. It was also shown to be superior to hybrid functionals to reproduce the dynamics and structural distortions in hybrid perovskites (*38*).

Density Functional Perturbation Theory (*39-41*) was used to compute the dielectric properties (dielectric constant and Born effective charges). In one case (BaTiO$_3$), the scissors correction was computed employing the HSEO6 hybrid functional (*42*). The structures were relaxed with 550 eV planewaves energy cutoff until residual forces were less than $10^{-5}$ eV/Å. The k-points grids were 8x8x8 (hybrid functional) to 16x16x16 (non-hybrid functional) for oxide and halide perovskites, and 4x2x4 for the hybrid MAPI. Spin orbit corrections were included in selected calculations. We tested the inclusion of SOC on the MAPI hybrid perovskites, reaching the conclusion that SCAN + SOC lowers the effective masses, particularly for the electrons, when compared to SCAN without SOC ($m_h^*$ = (0.18;0.12;0.18) vs (0.24;0.15;0.24) and $m_e^*$ = (0.14;0.08;0.14) vs (1.00;0.06;0.93)).

To compute the electrons shared and transferred, Bader basins (*43*) were defined using the Yu-Trinkle algorithm (*44*) and the corresponding localization (LI($\Omega$)) and delocalization indices ($\delta$ ($\Omega_1, \Omega_2$)) (*45*) were determined using the DGRID code (*46*). The initial wavefunctions are computed with ABINIT and the PBE-GGA exchange-correlation (*47*). The LIs are obtained by double integration of the same spin part of exchange-correlation hole over the corresponding basin, while DIs are obtained by integration of the same quantity once over the native basin and once over another basin. LI shows how many electron pairs are localized within the basin and do not participate in bonding, while DI provides the number of shared pairs of electrons between two atoms. The electron population of each atom N($\Omega$) is defined as the sum over the localization and half of the delocalization indices of an atom with all other basins. The same is obtained by simple integration of the electron density over the basin $\Omega$. The number of electrons transferred is obtained by subtracting the atomic charge from N($\Omega$). The values obtained for electron transfer (ET) and sharing (ES) are barely affected by the choice of the functional. This is no surprise, since these quantities only depend on the valence wave function. Furthermore, in the pair density calculation some of the missing exchange is reintroduced via the construction of a Slater determinant.

**Data availability**

**Acknowledgements**

We acknowledge the computational resources granted from RWTH Aachen University under project RWTH0508, as well as JARA0183 and JARA0198 in the initial project stage, kindly provided by Riccardo Mazzarello. This work was supported in part by the Deutsche Forschungsgemeinschaft (SFB 917), in part by the Federal Ministry of Education and Research (BMBF, Germany) in the project NEUROTEC (16ES1133 K) and in part by Excellence Initiative of the German federal and state governments (EXS-SF-neuroIC005). E. B. and J.-Y. R. acknowledge support from FNRS and computational resources provided by the CÉCI (funded by the F.R.S.-FNRS under Grant No. 2.5020.11) and the Tier-1 supercomputer of the Fédération Wallonie-Bruxelles (Walloon Region grant n°1117545). E. B. acknowledges the FNRS CDR project MULAN (J.0020.20). The improvement of several figures by Thomas Pössinger is gratefully acknowledged.

**Author contributions**

J.R. suggested a metavalent bonding mechanism in halide perovskites. M.W. initiated and conceptualized the project. J.-Y.R. and C.F.S. performed the tight-binding and DFT calculations, with early contributions from M.S, E.B. performed DFT calculations and provided insights on oxide perovskites, while P.G. helped with the DAFH analysis. The paper was written by M. W. with contributions from J.-Y.R. and C.-F.S. and support from all co-authors. All authors have given approval to the final version of the manuscript.

**Financial and conflict of interest**

The authors declare no conflict of interest.